\documentclass[aps,prb,twocolumn,superscriptaddress,groupedaddress]{revtex4}
\usepackage{graphicx}  
\usepackage{dcolumn}   
\usepackage{bm}        
\usepackage{amssymb}   
\usepackage[version=3]{mhchem} %Chemical Formulas
\usepackage{dsfont}
\usepackage{subfigure}
\usepackage[T1]{fontenc}

\DeclareMathAlphabet      {\mathbfit}{OML}{cmm}{b}{it}
\newcommand{\ket}[1]{\ensuremath{|#1\rangle}}
\newcommand{\bra}[1]{\ensuremath{\langle #1|}}

\newcommand{\diff}{\ensuremath{{\rm d}}}

\usepackage{color}

\begin{document}

\title{Pseudodoping of Metallic Two-Dimensional Materials by The Supporting Substrates}

\author{Bin Shao}
\affiliation{Institut f\"ur Theoretische Physik, Universit\"at Bremen, Otto-Hahn-Allee 1, 28359 Bremen, Germany}
\affiliation{Bremen Center for Computational Materials Science, Universit\"at Bremen, Am Fallturm 1a, 28359 Bremen, Germany}

\author{Andreas Eich}
\affiliation{Institute for Molecules and Materials, Radboud Univeristy, 6525 AJ Nijmegen, The Netherlands}

\author{Charlotte Sanders}
\affiliation{Department of Physics and Astronomy, Interdisciplinary Nanoscience Center (iNANO), Aarhus University, 8000 Aarhus C, Denmark}

\author{Arlette S. Ngankeu}
\affiliation{Department of Physics and Astronomy, Interdisciplinary Nanoscience Center (iNANO), Aarhus University, 8000 Aarhus C, Denmark}

\author{Marco Bianchi}
\affiliation{Department of Physics and Astronomy, Interdisciplinary Nanoscience Center (iNANO), Aarhus University, 8000 Aarhus C, Denmark}

\author{Philip Hofmann}
\affiliation{Department of Physics and Astronomy, Interdisciplinary Nanoscience Center (iNANO), Aarhus University, 8000 Aarhus C, Denmark}

\author{Alexander A. Khajetoorians}
\affiliation{Institute for Molecules and Materials, Radboud Univeristy, 6525 AJ Nijmegen, The Netherlands}

\author{Tim O. Wehling}
\affiliation{Institut f\"ur Theoretische Physik, Universit\"at Bremen, Otto-Hahn-Allee 1, 28359 Bremen, Germany}
\affiliation{Bremen Center for Computational Materials Science, Universit\"at Bremen, Am Fallturm 1a, 28359 Bremen, Germany}

\date{\today}

\begin{abstract}
We demonstrate how hybridization between a two-dimensional material and its substrate can lead to an apparent heavy doping, using the example of monolayer \ce{TaS2} grown on Au(111). Combining \textit{ab-initio} calculations, scanning tunneling spectroscopy experiments and a generic model, we show that strong changes in Fermi areas can arise with much smaller actual charge transfer. This mechanism, which we refer to as pseudodoping, is a generic effect for metallic two-dimensional materials which are either adsorbed to metallic substrates or embedded in vertical heterostructures. It explains the apparent heavy doping of \ce{TaS2} on Au(111) observed in photoemission spectroscopy and spectroscopic signatures in scanning tunneling spectroscopy. Pseudodoping is associated with non-linear energy-dependent shifts of electronic spectra, which our scanning tunneling spectroscopy experiments reveal for clean and defective \ce{TaS2} monolayer on Au(111). The influence of pseudodoping on the formation of charge ordered, magnetic, or superconducting states is analyzed. 

\end{abstract}

\maketitle

\section{Introduction}
%%%%%Strongy interacting 2d materials: which materials are out there. how are they produced; why they are cool: many-body + all-surface materials;
The family of two-dimensional (2d) materials has been expanding from graphene type materials to compounds such as 2d oxides, chalcogenides and halides\cite{novoselov_two-dimensional_2005,Novoselov_Review2016,2d_Atlas_Heine}. While most of the initial research concentrated on graphene and 2d semiconductors, monolayers of metallic 2d materials, particularly transition metal mono- and dichcalcogenides such as \ce{FeX} (X=Se, Te)\cite{ChinPhysLett2012,ge_superconductivity_2014,he_two-dimensional_2014,manna_interfacial_2017} and \ce{MX2} (M=V, Nb, Ta; X=S, Se)\cite{xi_strongly_2015,yu_gate-tunable_2015,cao_quality_2015,ugeda_characterization_2016,sanders_crystalline_2016-1}, can now be synthesized, processed and even integrated into van der Waals heterostructures\cite{Novoselov_Review2016}. A central reason for the growing interest in metallic 2d systems is that they host highly intriguing many-body states including competing superconducting, nematic, magnetic, excitonic, and charge ordered phases \cite{ChinPhysLett2012,ge_superconductivity_2014,he_two-dimensional_2014,manna_interfacial_2017,xi_strongly_2015,yu_gate-tunable_2015,cao_quality_2015,ugeda_characterization_2016}.

%%%%Doping
Generally, the electronic phase diagrams of these materials depend on carrier concentrations. Hence, one particularly important way of controlling many-electron phenomena in 2d materials is doping. It allows to switch, for instance, between insulating, charge-ordered, spin-ordered, or superconducting states of a material \cite{lee_doping_2006,taniguchi_electric-field-induced_2012,ye_superconducting_2012,yu_gate-tunable_2015,costanzo_gate-induced_2016}. There is however a problem in many systems: changes to electronic states (e.g. switching the prototypical material of 1T-\ce{TaS2} from a Mott insulator to a superconductor\cite{yu_gate-tunable_2015}) often require electron or hole doping on the order of a few 10$\%$ of an electron or hole per unit cell \cite{taniguchi_electric-field-induced_2012,ye_superconducting_2012,yu_gate-tunable_2015,costanzo_gate-induced_2016}. This translates into carrier concentrations $\gtrsim 10^{14}$\,cm$^{-2}$ which are out of reach for gating in standard field effect transistor geometries but require ionic liquids or chemical means like atom substitution, intercalation  etc. Doping at this level is potentially related to severe chemical changes of the material and substantial disorder \cite{Jeong1402}.

%In this paper, we discuss an alternative doping mechanism illustrated in Fig. \ref{fig:hyb_doping_scheme}. We show that hybridization of metallic 2d materials and substrates can provide effective carrier doping of more than 10$\%$ of an electron or hole per unit cell even in the case of weak physisorption. We explain how changes in Fermi volumes can arise without actual charge transfer between substrate and 2d material. We illustrate the order of magnitude of this effect with the real material example of \ce{TaS2} supported by Au and Pb metal surfaces. Afterwards the generic "pseudodoping" mechanism is explained within a simple two-band model and a self-energy formalism, which we use to discuss the impact of peudodoping on possible electronic phases. We argue that the recently observed strong doping of \ce{TaS2} on Au(111) \cite{sanders_crystalline_2016-1} is indeed largely hybridization induced.

In this letter, we discuss an alternative doping mechanism for metallic 2d materials. This mechanism, which we refer to as ``pseudodoping'', is associated with considerable changes in apparent Fermi areas but much smaller actual charge transfer between substrate and 2d material. The reason why a changed Fermi area does not necessarily imply charge transfer is that the Fermi contour is made up from hybridized states of the 2d material and the substrate. We illustrate that pseudodoping of more than 10$\%$ of an electron or hole per unit cell is possible using the example of monolayer 1H-\ce{TaS2} on Au(111), which has been recently studied by photoemission spectroscopy\cite{sanders_crystalline_2016-1}. Pseudodoping induced shifts of electronic states are often non-linearly energy-dependent such that they can affect states below and above the Fermi energy differently. In this way, pseudodoping gives a unified explanation of photoemission spectroscopy experiments\cite{sanders_crystalline_2016-1} and scanning tunneling spectra of \ce{TaS2} monolayers on Au(111). Afterwards the generic pseudodoping mechanism is explained within a simple two-band model, which we also use to discuss the influence of pseudodoping on the emergence of charge, magnetic, or superconducting order.

%\begin{figure}%
%\includegraphics[width=\columnwidth]{d00_Au_SF-TaS2.pdf}%
%\caption{(Color online) Coupling between a 2d material like \ce{TaS2} and a metal substrate. (a) Structure of \ce{TaS2} on Pb (111). Adsorption on the metal surface leads to hybridization $V$ between the electronic bands of the surface and the 2d material. As a consequence, there is an admixture of surface derived orbitals (gray) to the \ce{TaS2} states (purple) and level repulsion between the hybridizing bands illustrated in (b). The level repulsion leads to apparent hole-doping of the upper (surface derived) and electron doping of the lower (\ce{TaS2} derived) band.}%
%\label{fig:hyb_doping_scheme}%
%\end{figure}

\begin{figure*}%
	\includegraphics[width=\textwidth]{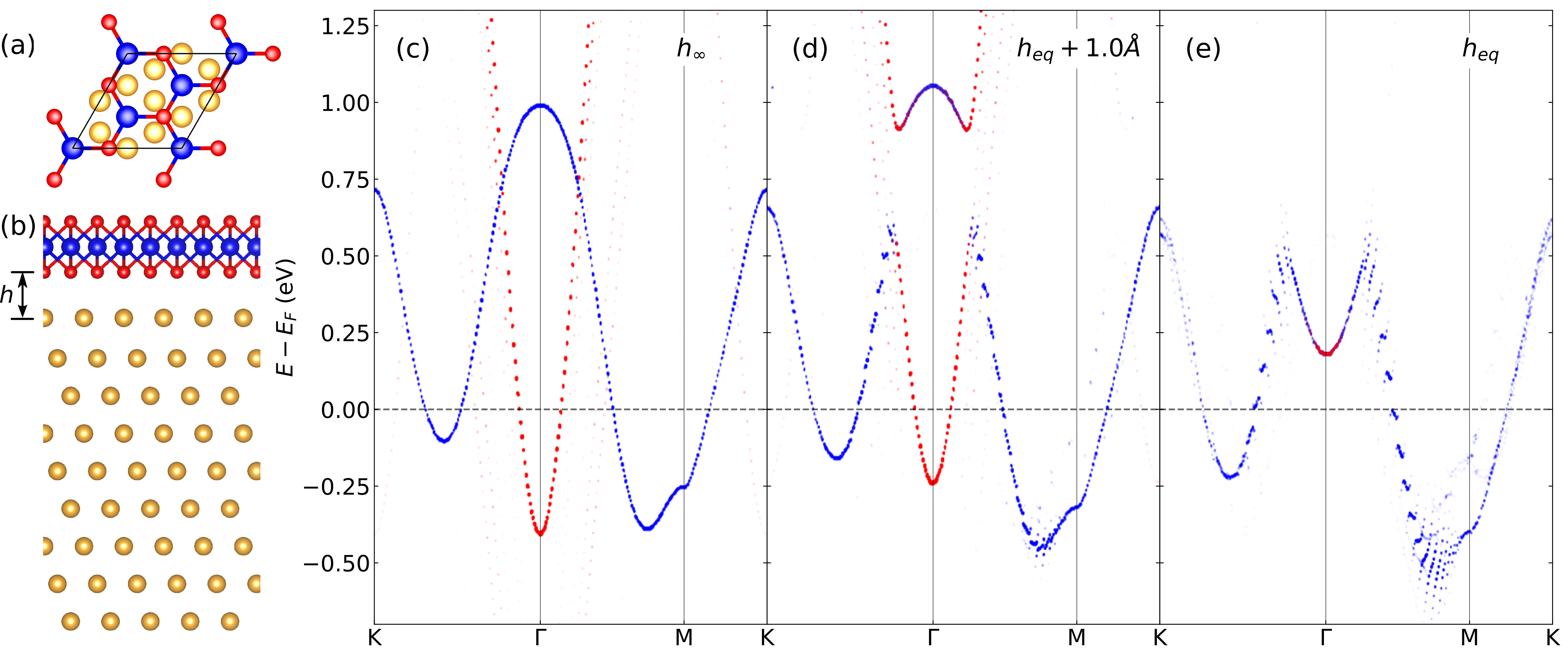}
	\caption{Supercell of monolayer 1H-\ce{TaS2} on Au(111), (a) top view and (b) side view. The adsorption height $h$ refers to the height of the lower S atoms above the upmost Au atoms. Projection of unfolded band structures on Ta-$d$ orbitals (blue dots) and upmost Au-$p_{\text{z}}$ orbitals (red dots) as a function of adsorption height: (c) $h_{\infty}$, (d) $h_{\text{eq}}+1$ \AA, and (e) $h_{\text{eq}}$. The $h_\infty$ case consists of two separated systems: a clean Au(111) substrate and a free-standing 1H-\ce{TaS2} monolayer.}%
	\label{fig:bands_structure}%
\end{figure*}

\section{Pseudodoping of \ce{TaS2} on Au(111)}
%\subsection{band shift and band separation} 

%We start with the example of a monolayer 1H-\ce{TaS2} adsorbed to the (111) surfaces of Au. %, which are widely used in surface science experiments. 
%From its interaction with graphene Au(111) is known as weakly coupling surfaces \cite{giovannetti_doping_2008,khomyakov_first-principles_2009,zhu_formation_2011}, i.e. the adsorption heights of graphene on these substrates exceed 3\AA\, which is indicative of van der Waals binding. 

To study the interaction of 1H-\ce{TaS2} monolayer and the Au substrate we performed density functional theory (DFT) calculations and scanning tunneling spectroscopy experiments. For the DFT simulations (see methods section for details), we constructed a $\sqrt{3}\times\sqrt{3}$\,R$30^\circ$ \ce{TaS2} supercell on a $2\times 2$ supercell of the Au(111) surface, where the Au surface has been laterally compressed by 0.5 $\%$ to have a commensurate structure. [See Fig.\ref{fig:bands_structure}(a) and (b).]
%The relaxed height $h$ of the lowest S atoms above the upmost Au atoms is 2.86~\AA, indicating a physicsorptive hybridization between \ce{TaS2} and the substrate.
The adsorption height of \ce{TaS2} above the Au(111) surface as denoted here by the vertical distance $h$ between the lowest S atoms and the upmost Au atoms has been optimized yielding the equilibrium distance $h_{\text{eq}}=2.86$\,\AA, which is indicative of physisorptive coupling between the \ce{TaS2} and the substrate.

%This height has been further raised from the equilibrium value $h_{\text{eq}}$ until the extreme case $h_{\infty}$, i.e. a clean Au(111) substrate and a free-standing \ce{TaS2} monolayer without interactions, so as 

The influence of the coupling between the \ce{TaS2} and the substrate on the electronic structure can be inferred from band structures in Fig.\ref{fig:bands_structure}(c) to (e), where the \ce{TaS2} layer approaches the Au(111) surface from $h_\infty$, i.e. the limit of a clean Au(111) substrate and a free-standing \ce{TaS2} monolayer, via {$h=h_{\rm eq}+1$\,\AA} to $h=h_{\rm eq}$. Unfolded band structures\cite{medeiros_effects_2014,medeiros_unfolding_2015} of the supercell to the Brillouin zone (BZ) of the \ce{TaS2} primitive cell are shown, where we highlight states with Ta-$d$ character as blue dots. It can be seen that the Ta-$d$ spectral weight of the adsorbed \ce{TaS2} at {$h_{\rm eq}+1$\,\AA} [Fig.\ref{fig:bands_structure}(d)] and $h_{\rm eq}$ [Fig.\ref{fig:bands_structure}(e)] roughly follows the energy dispersion of the free-standing \ce{TaS2} $d$-bands [Fig.\ref{fig:bands_structure}(c)] for most parts of the BZ path, particularly below the Fermi level. The $p_z$ spectral weight from the upmost Au atoms (red dots) results from a Shockley-type surface state with parabolic dispersion. However, there are also some prominent changes in the electronic dispersion of system, as \ce{TaS2} gradually approaches its substrate:

First, the Au-$p_\text{z}$ derived band at the $\Gamma$ point moves from $-0.4$~eV below the Fermi energy ($E_F$) to $0.2$~eV above the $E_F$ and thus gets depopulated for a decreasing adsorption height. Such energy shifts of surface states are quite typical for noble metal substrates\cite{forster_importance_2008,kowalczyk_investigation_2008}. In contrast, the Ta-$d$ derived band is lowered in energy by about 0.1~eV in the large parts of the BZ path for $h=h_{\rm eq}$ when compared to the free-standing case. This shift applies in particular to the \ce{Ta}-$d$ states at and below the Fermi level and translates via the density of states (DOS) $\rho(E_F)\approx 2$~eV$^{-1}$ into an apparent doping on the order of 0.2 electrons per \ce{TaS2} unit cell. This amount of doping appears unusual given that \ce{TaS2} is merely physisorbed on Au(111) but it is in agreement with the experiments: our calculated band structures at $h_{\text{eq}}$ [Fig.\ref{fig:bands_structure}(e)] well reproduce the occupied part of the energy dispersion measured by angle-resolved photoemission spectroscopy (ARPES) in Ref.~\onlinecite{sanders_crystalline_2016-1}. Summarizing, DFT and ARPES yield an apparent doping of $\gtrsim 0.2$ electrons per unit cell but the physical mechanism behind it is so far unclear.

A charge transfer $\Delta N$ (given in electrons per unit cell) over an effective distance $d$ is associated with an electrostatic potential difference $\Delta V=\alpha d\Delta N$, where $\alpha=e^2/\epsilon_0 A=18.6$ eV/\AA\, and $A=9.7$\,\AA$^2$ is the area of the \ce{TaS2} unit cell. Even if we assume that the effective distance between the charges in the 2d monolayer and the substrate is $d=1$\,\AA, i.e. much smaller than the typical distances of $\sim 4$\,\AA\, from the center of the \ce{TaS2} layer to the upmost substrate atoms, a charge transfer of 0.2$e^-$ 
%as found theoretically in the case of 1H-\ce{TaS2} on Au(111) 
would translate into a potential energy difference of $\Delta V\approx 4$~eV. Comparing the work functions of \ce{TaS2}, $W_{\rm TaS_2}=5.6$~eV \cite{shimada_work_1994} and Au(111), $W_{\rm Au}=5.31$~eV\cite{michaelson_work_1977},  it is clear that potential energy differences on the order of $\Delta V\approx 4$~eV are unexpected. Indeed, estimates for the charge transfer $\Delta N$ based on work function differences \cite{giovannetti_doping_2008,khomyakov_first-principles_2009} generally arrive at $\Delta N \ll 0.1e^-$ per unit cell for substrates like Au. 

We argue in the following that \textit{metallic} 2d materials are instead prone to ``pseudodoping'' with substrate induced changes in Fermi areas, which do \textit{not} primarily relate to charge transfer but rather to hybridization in a regime which is special to physisorbed 2d materials. To arrive at this conclusion, it is insightful to analyze also the unoccupied part of the electronic structure of the adsorbed \ce{TaS2}, Fig.\ref{fig:bands_structure}(c) to (e). In contrast to the almost rigid downward shift of the \ce{TaS2} derived bands below the $E_F$, we find strongly momentum and energy dependent changes in the electronic dispersion above the $E_F$. In particular, avoided crossings between the Au surface state and the \ce{TaS2} states strongly reshape the unoccupied part of the band structure near $\Gamma$, which indicates significant hybridization between the substrate and \ce{TaS2} layer.

\section{Scanning tunneling spectroscopy}
\begin{figure*}%
	\includegraphics[width=\textwidth]{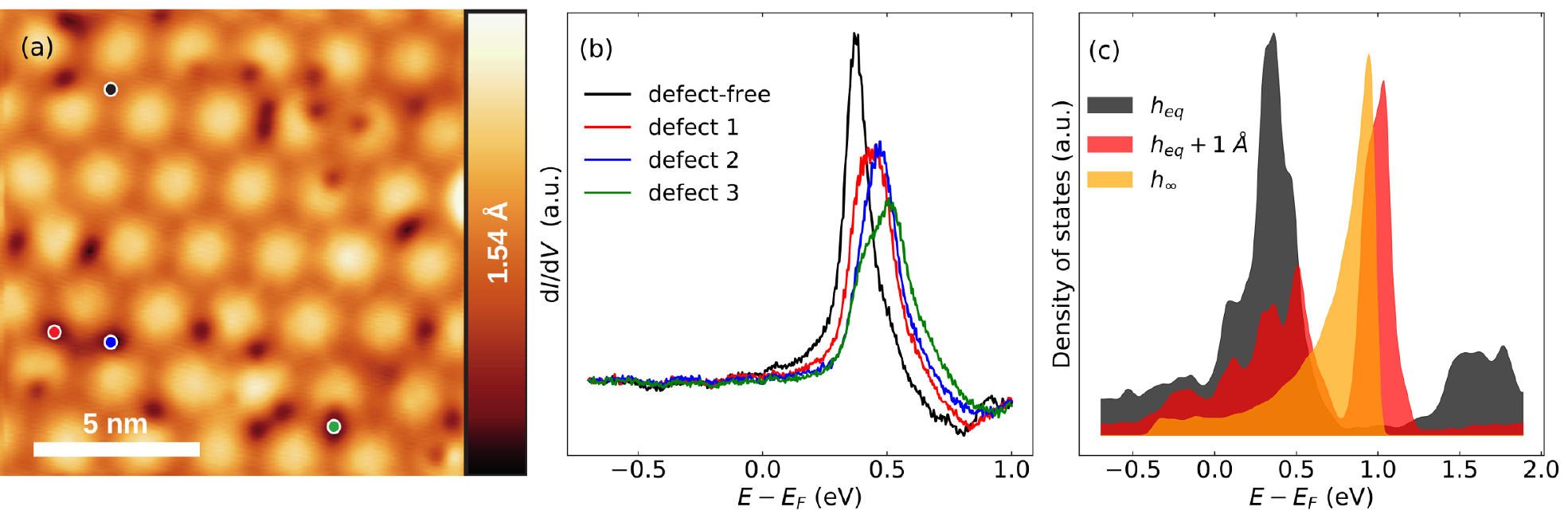}
	\caption{(a) High-resolution STM topography in the constant-current mode, where the color bar is related to the measured apparent height, STM parameters: $V_{\text{S}}=335$ mV, $I_{\text{T}}=500$ pA. (b) Comparison of STS point spectra from the sites with/without defects, STS parameters: $V_{\text{stab}}=1000$ mV, $I_{\text{stab}}=500$ pA, $V_{\text{mod}}=5$ mV. The corresponding measurement sites are represented by the spots in (a). (c) STS simulations as a function of the adsorption height $h$.}%
	\label{fig:sts_dos}%
\end{figure*}
%So far there has been no experimental study of these strong hybridization effects in the electronic structure above $E_F$. To change this situation, we performed scanning tunneling microscopy (STM) and spectroscopy (STS) experiments on 1H-\ce{TaS2} grown on Au(111).\cite{suppl} 
While ARPES confirms the apparent electron doping of the filled state bands, we utilize scanning tunneling microscopy/spectroscopy (STM/STS) to probe the empty state density of states.
The measurements were conducted with an Omicron LT-STM operated at 5K. The base pressure was better than $1\times10^{-10}$ mbar. An electrochemically etched tungsten wire, flashed \textit{in situ}, was used as the tip. The bias was applied to the sample. Topography images were taken in constant-current mode. Spectroscopy was performed utilizing a lock-in technique and the differential conductance was recorded with a modulation voltage of $V_{\text{mod}} = 5$ mV at a frequency of $f_{\text{mod}} = 5356$ Hz. The sample was prepared as discussed in a previous paper\cite{sanders_crystalline_2016-1}.
%The measurements were conducted on an Omicron LT-STM operated at 5 K, the bias was applied to the sample. The base pressure was better than $1\times10^{-10}$ mbar. An electrochemically etched tungsten wire, flashed \textit{in situ}, was used as the tip. Spectroscopy was performed utilizing a lock-in technique and recording the differential conductance with a modulation
%frequency of $f_{\text{mod}} = 5356$ Hz and a typical modulation voltage of $V_{\text{mod}}=5$ mV. Topography images were taken in constant-current mode. The sample was prepared as discussed in a previous paper\cite{sanders_crystalline_2016-1}. 
%The statements in the supplemental material: STM/STS experiments were performed on a commercial low-temperature STM operating at 5K, with the bias applied to the sample. The base pressure was better than $1\times10^{-10}$ mbar. An electrochemically etched tungsten wire, flashed \textit{in situ}, was used as the tip. Spectroscopy was performed utilizing a lock-in technique and recording the differential conductance, with a modulation frequency of $f_{\text{mod}} = 5356$ Hz. Topography images were taken in constant-current mode.
Fig~\ref{fig:sts_dos}(a) shows the morphology of 1H-\ce{TaS2} measured by STM.
%, in which the black areas refer to certain point defects. 
%Scanning tunelling spectra [Fig.~\ref{fig:sts_dos}(b)] were taken at the corner site of hexagonal rings with/without defects, as represented by the spots in Fig.~\ref{fig:sts_dos}(a). \tw{what are the hexagons; explanation of moire}
STS was taken at various points within the moir\'e unit cell of the surface, near and far away from apparent point defects. Previous studies of point defects in \ce{MoS2}/Au(111) indicate that such point defects arise from layer-dependent sulfur vacancies within the film\cite{krane_electronic_2016,krane_moire_2018}.
%\tw{Broken refs. Bin, please take care of them.}
We observe a non-dispersing state at 0.4 eV, which shifts in proximity of various defects.

To interpret these spectra, it is insightful to compare them to DFT-simulated STS [Fig.~\ref{fig:sts_dos}(c)], where we calculated the DOS inside a sphere at 5.0 \AA\ above the upmost S atoms of \ce{TaS2} for different seperations $h$ between the \ce{TaS2} and the Au(111) substrate. As shown in Fig.~\ref{fig:sts_dos}(c), for $h_{\infty}$ the main peak appears near 0.9 eV above the $E_F$. As $h$ decreases, the peak is split into two parts due to the increasing hybridization between the Ta-$d$ band and the Au(111) surface state. I.e. the peak splitting in the simulated STS directly reflects the avoided crossing between the two aforementioned bands visible in Fig. \ref{fig:bands_structure} (c)-(e). The lower part of the split peak in the simulated STS eventually shifts to about 0.4 eV above the $E_F$ for $h=h_{\rm eq}$, which is in line with the STS spectrum at the site without defects [Fig.~\ref{fig:sts_dos}(b)]. Indeed, the simulated spectrum at $h_{eq}$ well accounts for the measured spectrum at the site without defects, while the calculated spectrum for the free-standing case ($h_{\infty}$) deviates clearly from this STS spectrum. These findings in the defect-free case represent a first experimental observation of the theoretically predicted hybridization effect.

%Moreover, we have experimentally studied the influence of the defects in 1H-\ce{TaS2}@Au(111) on the electronic spectra. The defects 
Additional information about the interaction between the \ce{TaS2} and the substrate can be obtained by inspecting STS spectra collected in the vicinity of structural defects. Those spectra are also shown in Fig.~\ref{fig:sts_dos}(b). The presence of the defects
shifts the peak in STS spectrum to higher energies and reduces its amplitude with respect to the spectrum without defects [Fig.~\ref{fig:sts_dos}(b)]. Interestingly, those defect-induced changes are similar to the modifications that we find in the calculated spectra, when increasing the adsorption height from $h_{eq}$ to $h_{eq}+1$\AA\,[Fig.~\ref{fig:sts_dos}(c)]. Although the exact atomistic nature of the defects in the experiment is unknown, it appears that the defects can induce a local potential variation and 
%partially break the lateral ordering and 
change the vertical coupling strength of the interface. This can induce an energy shift of the surface state of Au substrate\cite{kowalczyk_investigation_2008}, 
which is qualitatively similar to the effect of changing adsorption heights. It is thus understandable that the defect-induced changes in the measured tunneling spectra [Fig.~\ref{fig:sts_dos}(b)] are very similar to the simulated spectra [Fig.~\ref{fig:sts_dos}(c)] with variations in the adsorption height $h$ and resulting changes in the hybridization strength. 

Taken together, the STS experiments confirm the significant impact of the hybridization on the electronic structure of the adsorbed \ce{TaS2}. We argue in the following that this hybridization provides the key to understanding the apparent heavy doping of the \ce{TaS2} monolayer on the Au(111) substrate in such a way that the \textit{nominal} occupancy of the \ce{TaS2} states varies considerably, while keeping the actual amount of charge in \ce{TaS2} orbitals almost the same as in the free-standing layer.

%Particularly in the unoccupied part of the spectrum there is a strongly energy dependent reshaping of the spectra, which is markedly different from the mostly rigid shift of the \ce{TaS2} band below $E_F$.

%Meanwhile, STS proofs itself being of a remarkable way to detect the pseudodoping effect.
%
%that the hybridization with substrates plays a essential role on understanding of 2d material electronic properties.
%Thus, STS is an ideal tool to reveal the effect of hybridization between 2d materials and substrates on electronic properties.
%Thus, it is clear that the hybridization plays a crucial role  

\section{Minimal model of pseudodoping}

%\tw{Bin, please rename the hybridization from $V$ to $t_\perp$ in order to not have $V$ used with two different meanings. Also take care of Fig 3 correspondingly.}
%\subsection{Two-band model}
%
%To address the physical scenario, we consider a model involving three bands derived from orbitals of the 2d material, the surface state and bulk states of the substrate, where we assume two hybridization terms $V_{SF}$ and $V_{bulk}$. Here, we ignore the k-dependence of the hybridization. The resulting Hamiltonian reads
%
% Here we consider 
%
%\begin{equation}
%	\hat{H_k}=
%	\begin{pmatrix}
%	\epsilon_{2d}(k)  &  V_{SF} & V_{bulk}     \\
%	V_{SF}   & \epsilon_{SF}(k)  & 0     \\
%	V_{bulk}   & 0   & \epsilon_{bulk}(k)   \\
%	\end{pmatrix}.
%\end{equation}
%The strength of the confining potentials
%%
%%The laterally ordering is not destroyed and the surface state is preserved. However, the surface potential is changed, leading to an energy shift of the surface state energy.
%%
%The Shockley surface states supported by the Au(111) is given by,
%
%\begin{equation}
%	\epsilon_{SF}(\vec{k_{\parallel}}) =E_{F} -E^{SF}_0(h) + {\hbar^2 k_{\parallel}^2}/{2m^{\star}}	
%\end{equation}
%\begin{equation}
%\epsilon_{bulk}(\vec{k_{\parallel}}) =E_{F} -E^{bulk}_0 + {\hbar^2 k_{\parallel}^2}/{2m}	
%\end{equation}
%
%Any modification of the surface barrier potential, e.g. temperature, surface contamination, defects, will consequently also vary the surface state energy potential
\begin{figure}%
	\includegraphics[width=\columnwidth]{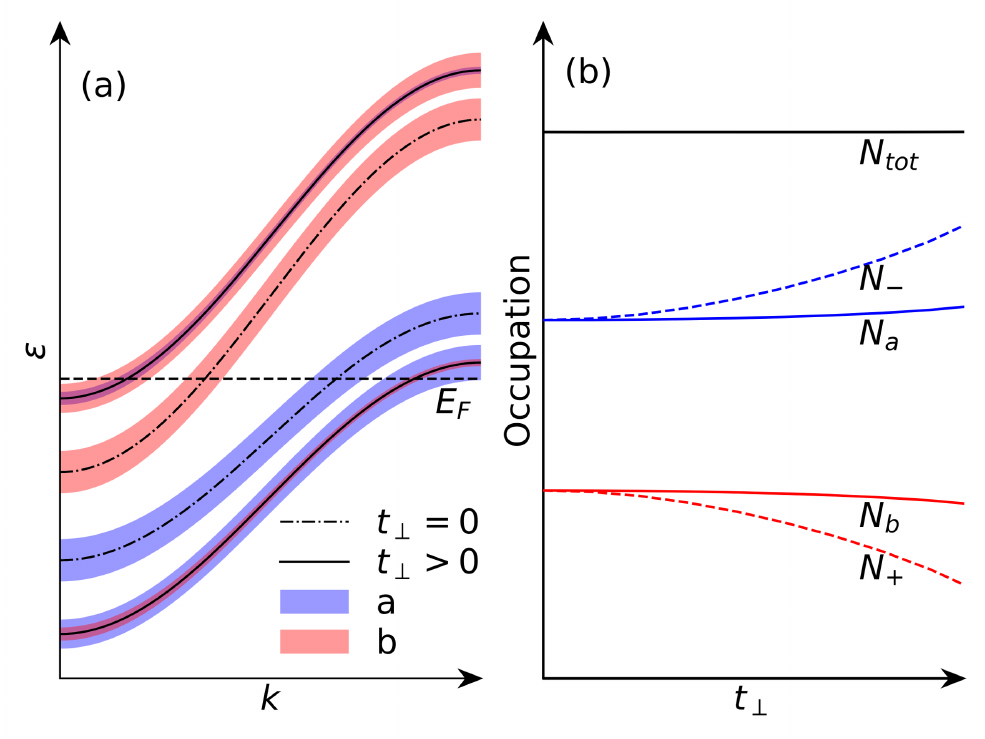}
	\caption{Evolution of (a) the energy dispersion and (b) the orbital/band occupation ($N_{a/b}$/$N_{\mp}$) as a function of the hybridization ($t_\perp$) between the orbital \ket{a} of 2d materials and \ket{b} of substrates. As illustrated in (a), the hybridization leads to an admixture of substrate derived states (red) to the 2d material states (blue) and a level repulsion between the hybridizing bands. The level repulsion, as $t_\perp$ increases, allows the band occupation $N_{\mp}$ to vary sizably, i.e. the upper band \ket{+} (the lower band \ket{-}) is apparently doped by holes (electrons), while leaving the orbital occupation $N_{a/b}$ almost unchanged as shown in (b). $N_{\text{tot}}$ refers to the total occupation number of the system, which is a constant under the variation in $t_\perp$.}%
	\label{fig:model}%
\end{figure}

%\textbf{actual occupancy and nominal occupancy}
%
%\textbf{actual doping and apparent doping}

We consider a minimal model involving two bands derived from the orbital $\ket{a}=(1,0)$ of the 2d material and $\ket{b}=(0,1)$ of substrate, where we assume a (real valued) hybridization $t_\perp\geq 0$ and a constant offset $-2\Delta$ in the on-site energies between the two. The resulting Hamiltonian reads
\begin{equation}
\hat H(k)=\epsilon_0(k)\mathbf{1}+t_\perp\sigma_1-\Delta\sigma_3,
\label{eq:H_2b2}
\end{equation}
where we used the Pauli matrices $\sigma_i$ and summarized all $k$-dependence of the initial dispersion in $\epsilon_0(k)$. The eigenstates of the full Hamiltonian are 
\begin{align}
\ket{-}&=\cos(\varphi)\ket{a}-\sin(\varphi)\ket{b},\nonumber\\
\ket{+}&=\cos(\varphi)\ket{b}+\sin(\varphi)\ket{a}
\end{align}
with $\tan2\varphi=t_\perp/\Delta$ and corresponding energies 
\begin{equation}
\epsilon_{\mp}(k)=\epsilon_0(k)\mp\sqrt{\Delta^2+t_{\perp}^2}.
\label{eq:eps_pm}
\end{equation}
As shown in Fig.~\ref{fig:model}(a), the hybridization yields an admixture of substrate derived states to the 2d material states and vice versa as well as a level repulsion between the hybridizing bands.

With $E_F=0$, the ground state density matrix reads
\begin{equation}
\hat N(k)=\Theta(-\epsilon_-(k))\ket{-}\bra{-}+\Theta(-\epsilon_+(k))\ket{+}\bra{+}.
\end{equation}

The central point is now to contrast the occupation of the bands 
\begin{equation}
N_\mp=\int\diff^2 k \bra{\mp}\hat N(k)\ket{\mp}
\end{equation} with the occupation of the orbitals 
\begin{equation}
N_{i}=\int\diff^2 k \bra{i}\hat N(k)\ket{i},\text{where}\; i\in\{a,b\}.
\end{equation}
The `nominal' band occupations $N_\mp$ are those measured in ARPES. In contrast, the orbital occupations $N_{a}$ and $N_{b}$ correspond to the actual charges in the 2d layer and the substrate, respectively, manifest in core level spectroscopies and also determine electric fields at the interface. Here, we have $N_a=\cos^2(\varphi)N_-+\sin^2(\varphi)N_+$ and $N_b=\cos^2(\varphi)N_++\sin^2(\varphi)N_-$.

For vanishing hybridization ($t_\perp=0 \Rightarrow \varphi=0$) the orbital occupancies simply coincide with the band occupancies: \[N^0_{a/b}=N^0_{\mp} =\int_{-\infty}^0\diff \epsilon\rho^0(\epsilon\pm\Delta),\] where $\rho^0(\epsilon)=\int{\diff^2 k}\delta(\epsilon-\epsilon_0(k))$ is the DOS associated with the dispersion $\epsilon_0(k)$. At finite $t_\perp$ the band fillings become 
\begin{align}
N_\mp&=\int_{-\infty}^0\diff \epsilon\rho^0(\epsilon\pm\sqrt{\Delta^2+t_{\perp}^2})\nonumber\\
&=N^0_\mp+\int_{-\infty}^0\diff \epsilon[\rho^0(\epsilon\pm\sqrt{\Delta^2+t_{\perp}^2})-\rho^0(\epsilon\pm\Delta)]\nonumber\\
%&\approx N^0_\mp\pm\int_{-\infty}^0\diff \epsilon (\partial_\epsilon \rho^0(\epsilon))(\sqrt{\Delta^2+t_{\perp}^2})-\Delta)\nonumber\\
&\approx N^0_\mp\pm \rho^0(0)t_{\perp}^2/2\Delta
\end{align}
I.e. in ARPES experiments, the upper (lower) band appears hole (electron) doped by an amount of $\rho^0(0)t_{\perp}^2/2\Delta$ [c.f. dashed lines in Fig.~\ref{fig:model}(b)]. 

However, the occupancy of the orbital $\ket{a}$ in the 2d layer is
\begin{align}
N_a&=\cos^2(\varphi)N_- + \sin^2 (\varphi)N_+\nonumber\\
%&=N_-+\sin^2 (\varphi) (N_+-N_-)\nonumber\\
%&\approx N_-+ \varphi^2 (N^0_+-N^0_-)\nonumber\\
&\approx N^0_-+\rho^0(0)\left[t_{\perp}^2/(2\Delta)+ \varphi^2 (-2\Delta)\right]\nonumber \\
%&\approx N^0_-+\rho(0)(t_{\perp}^2/2\Delta+ (t_{\perp}/2\Delta)^2 (-2\Delta)\nonumber \\
&= N^0_- + O(t_{\perp}^3).
\end{align}
Analogously, we find $N_b\approx N^0_+$ to second order in $t_{\perp}$ [c.f. solid lines in Fig.~\ref{fig:model}(b)]. I.e. the actual charge transfer cancels to leading order in $t_{\perp}$ while the `nominal' doping does not. This is the key point of the pseudodoping mechanism. The hybridization, as depicted in Fig.~\ref{fig:model}, allows the occupancies $N_-$ of the lower band mainly contributed by the 2d material states to vary considerably, while keeping the actual amount of charge $N_a$ in the 2d material almost the same as in its free-standing form.

\section{Pseudodoping and electronic instabilities}
Interaction terms like Coulomb interactions naturally couple localized states. In the example of a Hubbard type interaction they are of the form
\begin{equation}
H_U=U\sum_{i} \hat n^a_{i\uparrow} \hat n^a_{i\downarrow},
\label{eq:H_int}
\end{equation}
where $i$ refers to the lattice site of the electrons in the 2d layer, $\sigma=\{\uparrow,\downarrow\}$ to their spin and $\hat n^a_{i\sigma}$ are the corresponding occupation number operators. $U$ is the on-site interaction matrix element and can describe repulsive ($U>0$) or attractive interaction ($U<0$). 

We assume that the substrate states are non-interacting and that their DOS at the Fermi level is small as compared to the DOS of the 2d material. Then, the admixture of the substrate states to the bands of the 2d material upon hybridization reduces the \textit{effective} interaction inside the hybridized band by a factor of $Z^2=\cos^4(\phi)$, i.e. $U\to U^{\rm eff}=Z^2 U$.

\begin{figure}
	\includegraphics[width=\columnwidth]{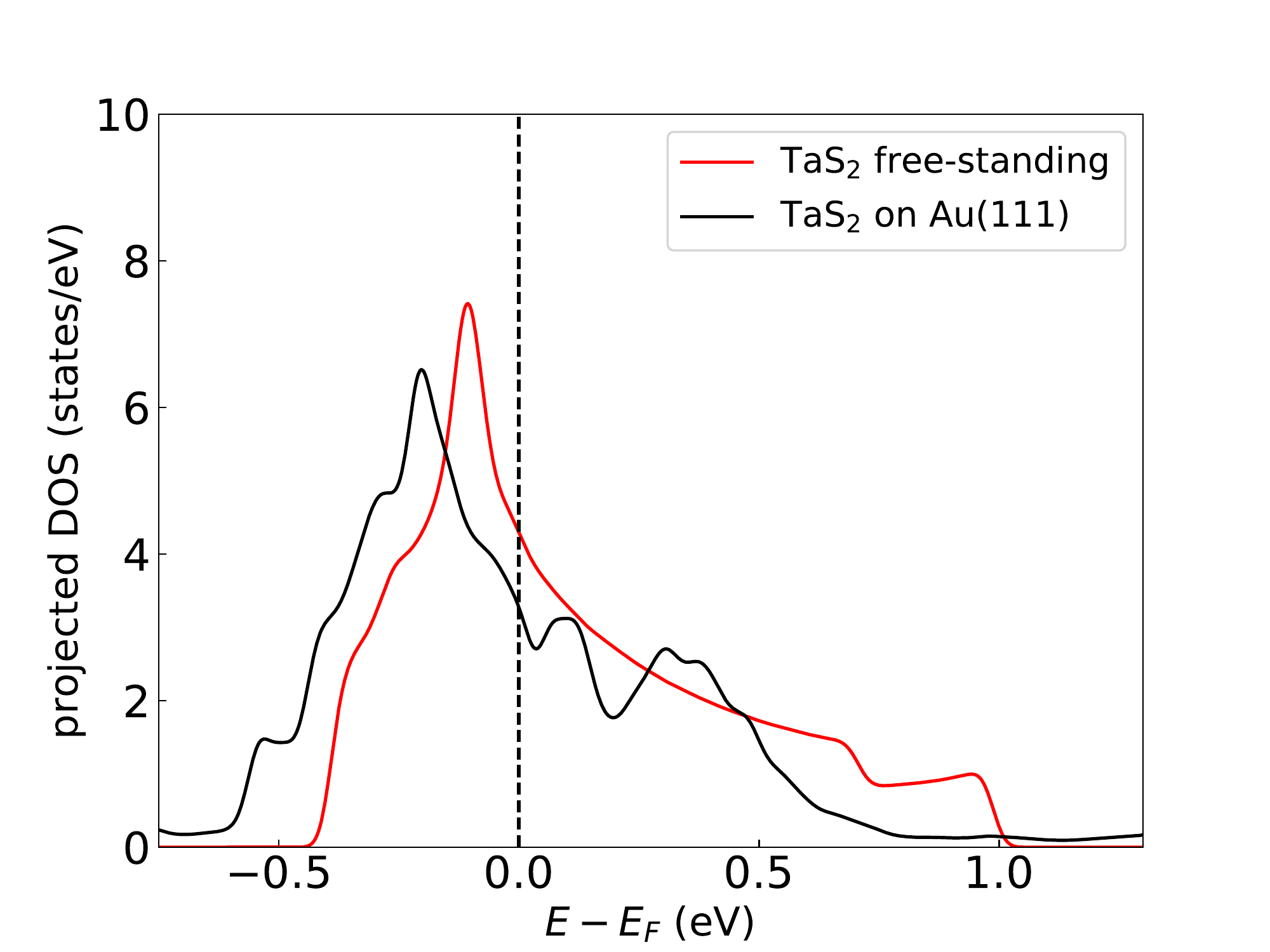}
	\caption{Comparison of the DOS projected on Ta-\textit{d} orbitals of 1H-\ce{TaS2} on Au(111) substrate (black line) and in its free-standing form (red line). }
%	\tw{Bin: Please specify here what is shown excatly. I.e. total DOS or Ta d-projected DOS. In my view only the latter makes sense on the substrate...}}%
	\label{fig:dos_comparison}%
\end{figure}

For weak coupling instabilities such as BCS superconductivity, characteristic transition temperatures $T_C\sim\exp[-1/|U^{\rm eff}\rho|]$ are determined by the interaction $U^{\rm eff}$ and DOS at the $E_F$ $\rho$ resulting from the hybridized band. If the DOS of the original 2d band is structureless, we expect simply a reduction of $T_C$. An analogous line of argumentation holds for weak coupling charge- or spin-density wave instabilities at some wave vector $Q$, where the susceptibility $\chi(Q)=\int\diff^2 k (f(\epsilon_k)-f(\epsilon_{k+Q}))/(\epsilon_k-\epsilon_{k+Q})$ plays the role of an effective density of states. Thus, hybridization effects should quench tendencies towards weak coupling electronic instabilities, if the DOS / generalized susceptibilities are essentially independent of the $E_F$. An even stronger suppression of instabilities is expected if $\rho$ or $\chi(Q)$ are reduced upon hybridization related pseudodoping. 
A comparison of the DOS of 1H-TaS$_2$ on Au(111) and in its free-standing form (c.f. Fig.~\ref{fig:dos_comparison}) shows that there is indeed a reduction of $\rho(E_F)$ upon deposition on Au, i.e. both, the reduction of effective coupling constants and the reduced DOS\cite{Calandra_PRB}, will contribute here to the suppression of charge density wave / superconducting states observed in Ref. \onlinecite{sanders_crystalline_2016-1}.

If, however, hybridization shifts a strong peak in the DOS of the 2d material's band towards $E_F$ such that the hybridization induced increase in $\rho$ overcompensates the reduction of the effective interaction by the factor $Z^2$, we arrive at an increase in $T_C$. Additionally, if there is a sizable interaction between electrons in the substrate states, the effective interaction can indeed be enhanced upon hybridization with the substrate, which finally also yields an increase in $T_C$. Such an `inherited' interaction from the substrate has been conjectured to contribute to the strongly enhanced superconducting critical temperature of FeSe when coupled to \ce{SrTiO3}\cite{lee_interfacial_2014}.

\section{Conclusions} 
In summary, we have introduced the mechanism of ``pseudodoping'', which can lead to apparent heavy doping of 2d materials on metallic substrates even in case of weak physisorption. Pseudodoping manifests itself as surprisingly large shifts of bands in ARPES experiments, as measured for \ce{TaS2} on Au(111) \cite{sanders_crystalline_2016-1}. Yet, pseudodoping is expected to be ubiquitous in metallic 2d materials grown on 2d metallic substrates. It is due to hybridization of metallic 2d materials with electronic state from the supporting substrates and involves much less actual charge transfer between the layer and its substrate than changes in the Fermi areas would suggest. Nonetheless, instabilities of the electronic system towards symmetry broken states are expected to be highly sensitive to this kind of doping. Particularly, the prospect of utilizing pseudodoping to control electronic phase diagrams deserves future explorations.

%We have shown that hybridization of metallic 2d materials with their substrates can lead to apparent doping manifesting as shifts of bands in ARPES experiments. This "pseudodoping" can involve much less actual charge transfer between the layer and its substrate than changes in the Fermi surface would suggest. Nonetheless, instabilities of the electronic system towards symmetry broken states, particularly weak coupling instabilities, are expected to be highly sensitive to this kind of doping. The pseudodoping mechanism outlined, here, might explains the apparent strong doping of \ce{TaS2} on Au(111) \cite{sanders_crystalline_2016-1}.  Particularly, the prospect of utilizing pseudodoping to control electronic phase diagrams deserves future explorations.

%%%BCS M~Z + changes in rho, where rho follows Delta N, i.e. pseudodoping affects rho as entering BCS theory. 
%%%Slater Antiferromagnetism: Same phenomenology holds. Diverging chi(Q).
%%%Strong coupling: Situation more complicated: Local moment formation essentially unaffected hybridization doping. Mott insulator cannot form. metallic substrate states will penerate into 2d layer via Kondo effect. 

\textit{Acknowledgments}
This work was supported by the European Graphene Flagship, by the Danish Council for Independent Research, Natural Sciences under the Sapere Aude program (Grant No. DFF-4002-00029) and by VILLUM FONDEN via the Centre of Excellence for Dirac Materials (Grant No. 11744). We acknowledge financial support from The Netherlands Organization for Scientific Research (NWO) via the VIDI project: `Manipulating the interplay between superconductivity and chiral magnetism at the single atom level' with project number 680-47-534. The numerical computations were carried out on the Norddeutscher Verbund zur F\"orderung des Hoch- und H\"ochstleistungsrechnens (HLRN) cluster.

%The unfolding has been performed using the BandUP code. 

%\appendix
\section*{Methods: Computational Details}
\label{app:methods}
We performed density functional theory calculations using the Vienna Ab Initio Simulation Package (VASP) \cite{kresse_norm-conserving_1994} with the projector augmented wave basis sets \cite{blochl_projector_1994,kresse_ultrasoft_1999} and the generalized gradient approximation to the exchange correlation potential \cite{perdew_generalized_1996}. In all cases we fixed the in-plane lattice constant of 1H-\ce{TaS2} to the experimental value of $a=3.316$\AA\; \cite{mattheiss_band_1973}. 

The Au(111) surfaces were modeled using slabs with thickness of 30 atomic layers and a single layer of 1H-\ce{TaS2} absorbed on the upper side of the slabs [see Fig. 1(a) and (b) in the main text]. The Au(111) slab was furthermore terminated with H atoms on the bottom side of the slab, i.e. on the side without \ce{TaS2} coverage. The lateral coordinates of all atoms were kept fixed and we laterally compressed the Au(111) surface by $0.5\%$ to match a ($\sqrt{3}\times\sqrt{3}$)R$30^\circ$ unit cell of 1H-\ce{TaS2} with a $2\times 2$ unit cell of the metal surface. The vertical coordinates of the Ta and S atoms were relaxed until forces acting on them were below {0.01~eV/\AA} leading to structures, where the closest vertical distance between S atoms and Au surface atoms is 2.86~\AA.

%\bibliography{new_ref}
%\bibliography{Hyb_doping}
%\bibliographystyle{apsrev}

\end{document}